\documentstyle[prl,aps,multicol,array,psfig]{revtex} 
\begin{document} 
 
\draft 
 
\title{'t Hooft-Polyakov Monopoles in an Antiferromagnetic Bose-Einstein 
Condensate} 
 
\author{H.T.C. Stoof, E. Vliegen, and U. Al Khawaja} 
\address{Institute for Theoretical Physics, 
         University of Utrecht, Princetonplein 5, \\ 
         3584 CC Utrecht, 
         The Netherlands} 
 
\maketitle 
 
\begin{abstract} 
We show that an antiferromagnetic spin-1 Bose-Einstein condensate, which can for 
instance be created with $^{23}$Na atoms in an optical trap, has not only 
singular line-like vortex 
excitations, but also allows for singular point-like 
topological 
excitations, i.e., 't Hooft-Polyakov monopoles. We discuss the 
static and dynamic properties of 
these monopoles. 
\end{abstract} 
 
 
\begin{multicols}{2} 
{\it Introduction.} --- Quantum magnetism plays an important role in such 
diverse areas of physics as high-temperature superconductivity, quantum phase 
transitions and the quantum 
Hall effect. Moreover, it now appears that magnetic 
properties are also very 
important in another area, namely Bose-Einstein 
condensation in trapped atomic 
gases. This is due to two independent 
experimental 
developments. The first development is the realization of an 
optical trap for 
$^{23}$Na atoms \cite{MIT1}, whose operation no longer requires 
the gas to be doubly spin-polarized and has given rise to the creation of a 
spin-1 Bose-Einstein antiferromagnet \cite{MIT2}. The second development is the 
creation of a two-component 
condensate of $^{87}$Rb atoms \cite{JILA1}, which by 
means of rf-fields can be manipulated so as to 
make the two components 
essentially equivalent \cite{JILA2}. As a result also a 
spin-1/2 Bose-Einstein 
ferromagnet can now be studied in detail experimentally. 
 
The spin structure of these condensates has recently been 
worked out by a number of authors \cite{jason1,OM,nick,jason2} and also the first studies of the line-like vortex excitations have appeared \cite{jason1,Yip,TK}. An immediate 
question that comes to mind, however, is whether the spin degrees of 
freedom lead also to other topological excitations that do not have an analogy 
in the case of a single component Bose-Einstein condensate. The answer to this 
question is in general affirmative. Indeed, we have recently shown that 
ferromagnetic Bose-Einstein condensates have long-lived 
skyrmion excitations, 
which are nonsingular but topologically nontrivial point-like spin textures 
\cite{usama}. Moreover, we show here that also spin-1 Bose-Einstein 
antiferromagnets have point-like topological excitations. In particular, there 
exist singular point-like spin textures, which we call 't Hooft-Polyakov 
monopoles because of their analogy with 
magnetic monopoles in particle physics \cite{mono}. Having done so, we then turn to the investigation of the 
precise texture and the dynamics of these monopoles. 
 
{\it Topological considerations.} --- To find the topological 
excitations of a spin-1 Bose-Einstein condensate, we need to know 
the full symmetry of the macroscopic wave function $\Psi({\bf r}) 
\equiv \sqrt{n({\bf r})} \zeta({\bf r})$, where $n({\bf r})$ is 
the total density of the gas, $\zeta({\bf r})$ is a normalized 
spinor that determines the average local spin by means of $\langle 
{\bf F} \rangle({\bf r}) 
    = \zeta^{\dagger}({\bf r}) {\bf F} \zeta({\bf r})$, 
and ${\bf F}$ are the usual spin matrices obeying the commutation relations 
$[F_{\alpha},F_{\beta}] 
                    = i \epsilon_{\alpha\beta\gamma} F_{\gamma}$. 
Note that here, and in the following, summation over repeated 
indices is always implicitly implied. From the work of Ho 
\cite{jason1} we know that in the antiferromagnetic case the mean-field 
interaction 
energy is minimized for 
$\langle {\bf F} \rangle({\bf r}) = {\bf 0}$, which implies that the parameter 
space for the spinor 
$\zeta({\bf r})$ is only $S^1 \times S^2$ because we are 
free to choose both its overall phase and the orientation of the spin 
quantization axis. Introducing the superfluid phase $\vartheta({\bf r})$ and the 
unit vector field ${\bf m}({\bf r})$, this topology can also be understood 
explicitly from the fact that all the spinors 
\begin{equation} 
\label{spinor} 
\zeta({\bf r}) = \frac{e^{i\vartheta({\bf r})}}{\sqrt{2}} 
\left( 
  \begin{array}{c} 
   - m_x({\bf r}) + i m_y({\bf r}) \\ 
   \sqrt{2} m_z({\bf r}) \\ 
   m_x({\bf r}) + i m_y({\bf r}) 
  \end{array} \right) \equiv e^{i\vartheta({\bf r})} \zeta_{\rm AF}({\bf r}) 
\end{equation} 
have a vanishing average spin and hence are locally degenerate. 
 
What does this tell us about the possible topological 
excitations? For line-like defects or vortices, we can assume 
$\zeta({\bf r})$ to be independent of one direction and the spinor 
represents a mapping from a two-dimensional plane into the order 
parameter space. If the vortex is singular this is visible on 
the boundary of the two-dimensional plane and we need to 
investigate the properties of a continuous mapping from a circle 
$S^1$ into the order parameter space $G$, i.e., of the first 
homotopy group $\pi_1(G)$ \cite{mermin}. Since $\pi_1(S^1 \times S^2) = Z$, 
we conclude that an antiferromagnetic spin-1 
condensate can have vortices with winding numbers that are an 
arbitrary integer. Physically, this means that by traversing the 
boundary of the plane, the spinor can wind around the order 
parameter an arbitrary number of times. 
 
Similarly we can also discuss singular point-like defects. Since the boundary of 
a three-dimensional gas is the surface of a three-dimensional sphere, 
singular point-like defects are determined by the second homotopy group 
$\pi_2(G)$ \cite{mermin}. Because also $\pi_2(S^1 \times S^2) = Z$, such 
topological excitations thus indeed exist in the case of a 
spin-1 Bose gas with 
antiferromagnetic interactions. In view of the work of 't Hooft and Polyakov we 
refer to these excitations 
as monopoles, although it 
would also be justifiable to 
call them singular skyrmions. In contrast to the nonsingular skyrmions in the 
Bose-Einstein ferromagnets, which inherently are nonequilibrium objects, the 't 
Hooft-Polyakov monopoles turn out to be thermodynamically stable excitations as 
we show next. 
 
{\it Monopole texture.} --- The grand-canonical energy of the spinor condensate 
can be obtained from the usual Gross-Pitaevskii theory, which for the restricted 
parameter space given in Eq.~(\ref{spinor}), leads to the expression 
\begin{eqnarray} 
\label{energy} 
E[n,\zeta] = \int d{\bf r}~ 
     \psi^*({\bf r}) 
       \left( - \frac{\hbar^2 \mbox{\boldmath $\nabla$}^2}{2m} 
              + V_{\rm trap}({\bf r}) - \mu 
       \right. \hspace*{0.3in} \nonumber \\ 
       \left. + \frac{g_n}{2} |\psi({\bf r})|^2 
              + \frac{\hbar^2}{2m} 
                [\mbox{\boldmath $\nabla$} {\bf m}({\bf r})]^2 
       \right) \psi({\bf r}) ~, 
\end{eqnarray} 
where $\psi({\bf r}) = \sqrt{n({\bf r})} e^{i\vartheta({\bf r})}$ is the 
superfluid order parameter, $V_{\rm trap}({\bf r}) = m \omega^2 {\bf r}^2/2$ is an isotropic harmonic trapping potential, $\mu$ is the chemical potential, and 
$g_n \equiv 4\pi a_n \hbar^2/m$ is the appropriate coupling constant for density 
fluctuations. Minimization of this energy determines both the spin texture 
${\bf m}({\bf r})$ and the density profile $n({\bf r})$ of the 't Hooft-Polyakov 
monopole \cite{spin}. Since gradients in the spin texture do not couple to the 
superfluid phase $\vartheta({\bf r})$, we can at this point already conclude 
that the presence of a monopole will not induce any superfluid flow in the 
atomic cloud. From now on we therefore no longer consider this degree of freedom.
 
Interestingly, the spin texture is uniquely determined by the fact that it 
should have a topological winding number 
\cite{R} 
\begin{equation} 
W = \frac{1}{8\pi} \int d{\bf r}~ \epsilon_{ijk} \epsilon_{\alpha\beta\gamma} 
      \partial_i m_{\alpha} \partial_j m_{\beta} 
      \partial_k m_{\gamma} 
\end{equation} 
equal to 1 and that it should also minimize the gradient energy 
\begin{equation} 
\label{gradient} 
E_{\rm grad}[n,\zeta] = \int d{\bf r}~ 
      \frac{n({\bf r}) \hbar^2}{2m} 
                  [\mbox{\boldmath $\nabla$} {\bf m}({\bf r})]^2~. 
\end{equation} 
As we discuss in more detail in a moment, the latter requires the spin texture 
to be as symmetric as possible. In combination with the first requirement, we 
thus conclude that ${\bf m}_{\rm HP}({\bf r}) = {\bf r}/r$ for a monopole in the 
center of the trap. Indeed, a spin texture with the same winding number can be 
obtained by rotating ${\bf m}_{\rm HP}({\bf r})$ by an arbitrary rotation matrix 
${\bf R}(r)$ that only depends on the radial distance $r$. As a result the 
gradient energy is turned into the suggestive form 
\begin{equation} 
E_{\rm grad}[n,\zeta] = \int d{\bf r}~ 
   \frac{n({\bf r}) \hbar^2}{2m} 
        |[\mbox{\boldmath $\nabla$} - i {\bf A}({\bf r})] 
                                             {\bf m}_{\rm HP}({\bf r})|^2~, 
\end{equation} 
which brings out even more clearly the analogy with the $O(3)$ gauge theory 
studied by 't Hooft and Polyakov. In our case the vector potential is, however, 
defined by ${\bf A}({\bf r}) 
      = i({\bf R}^{-1}(r) \mbox{\boldmath $\nabla$} {\bf R}(r))$ and is always 
orthogonal to the gradient of ${\bf m}_{\rm HP}({\bf r})$. The gradient energy 
is therefore minimized for ${\bf R}(r) = {\bf 1}$, as anticipated previously. 

\begin{figure}
\psfig{figure=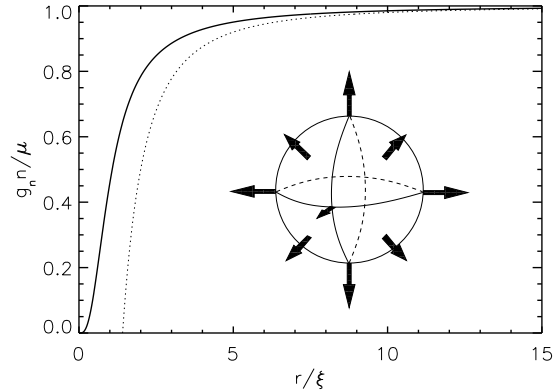,width=8cm,angle=-90}
\caption{\narrowtext Spin texture and density profile of the 
                     't Hooft-Polyakov monopole. The solid line shows the
                     exact numerical result, whereas the dashed line
                     shows the analytic long distance behaviour 
                     $n(r) \simeq (\mu/g_n) (1 - 2(\xi/r)^2)$.}
\end{figure} 

Substituting our hedgehog solution for the spin texture into 
Eq.~(\ref{energy}), we see that the density profile of the spinor 
condensate is found from a Gross-Pitaevskii equation with a 
centrifugal barrier equal to $\hbar^2/m r^2$. Considering first 
the homogeneous case and writing the superfluid order parameter as 
$\psi({\bf r}) = (\mu/g_n)^{1/2} f(r/\xi)$, with 
$\xi=(\hbar^2/2m\mu)^{1/2}$ the correlation length, we obtain 
explicitly that 
\begin{equation} 
\left( - \frac{1}{\rho^2} \frac{d}{d\rho} \left( \rho^2 \frac{d}{d\rho} \right)
       + \frac{2}{\rho^2} 
       + f^2(\rho) -1 \right) f(\rho) = 0~. 
\end{equation} 
This equation cannot be solved analytically, but its numerical 
solution is shown in Fig.~1. What is most important for our 
purposes, however, is the large distance behavior of the density 
profile. Neglecting the gradient terms in the left-hand side, we 
easily find that $f(\rho) \simeq 1 - 1/\rho^2$ for $\rho \gg 1$. 
Furthermore, the monopole is clearly seen to possess a core with a 
typical size of the order of the correlation length. Inside this 
core the density is strongly reduced, which offers the opportunity 
to detect monopoles by the same expansion experiments that have 
recently also been used to observe vortices \cite{jean}. 

Having obtained the spin texture and the density profile of the 
't Hooft-Polyakov monopole, we are now in a position to determine also its 
energy. Placing the monopole in the center of a spherical volume with a
large radius $R \gg \xi$, we obtain 
\begin{equation} 
\label{menergy} 
E_{\rm HP} = \frac{\mu}{a_n}(R - R_{\rm core})~, 
\end{equation} 
where $R_{\rm core}$ is the effective core size of the monopole. 
Its calculation requires knowledge of the complete density profile 
and we find numerically that $R_{\rm core} \simeq 1.4 \xi$. The monopole energy thus diverges linearly with the system size, which implies that in the thermodynamic limit only pairs of monopoles with opposite winding numbers require a finite energy for their creation.

The physical interpretation of Eq.~(\ref{menergy}) is that to 
calculate the monopole energy we only need to evaluate the 
gradient energy for a fixed density equal to $\mu/g_n$, but 
restrict the integration to the volume outside a spherical core 
region with radius $R_{\rm core}$. This interpretation is 
particularly useful for a trapped spinor condensate in the 
Thomas-Fermi limit, when the size $R_{\rm TF} = (2\mu/m\omega^2)^{1/2}$ 
of the condensate is much larger than the 
correlation length $\xi$. For a monopole at position ${\bf u}$ 
near the center of the trap, we find in this manner that 
\begin{equation} 
\label{pot} 
E_{\rm HP}[{\bf u}] \simeq \frac{2\mu}{3 a_n} R_{\rm TF} 
  \left( 1 - \frac{3 {\bf u}^2}{R_{\rm TF}^2} 
            + \frac{3 {\bf u}^4}{5 R_{\rm TF}^4} + \dots 
  \right)~, 
\end{equation}
if we neglect the core contributions that are smaller by a factor of 
$\xi/R_{\rm TF}$. The significance of this result will become clear once we understand the dynamical properties of the 't Hooft-Polyakov monopole. 
 
{\it Monopole dynamics.} --- In first instance we expect the dynamics of the 
monopole to be determined by the action 
$S[n,\zeta] = \int dt (T[n,\zeta] - E[n,\zeta])$ 
with a time-derivative term 
that is equal to 
\begin{equation} 
T[n,\zeta] = \int d{\bf r}~ 
  n({\bf r},t) 
     \zeta^{\dagger}({\bf r},t) 
                 i\hbar \frac{\partial}{\partial t} \zeta({\bf r},t) 
\end{equation} 
in the Gross-Pitaevskii theory. However, when we restrict 
ourselves to the antiferromagnetic spinor $\zeta_{AF}({\bf r},t)$, 
the time-derivative term in the action exactly vanishes due to the 
normalization condition ${\bf m}^2({\bf r},t) = 1$. To find any 
dynamics for the monopole we thus need to consider also 
fluctuations that bring the spinor condensate out of the 
antiferromagnetic order parameter space. In the Thomas-Fermi 
limit, the relevant dynamical part of the action thus becomes 
\begin{eqnarray} 
S_{\rm dyn}[n,\zeta] \hspace*{2.45in} \nonumber \\ = \int dt 
  \left( T[n,\zeta] 
     - \int d{\bf r}~ \frac{g_s}{2} [n({\bf r},t) 
                 \langle {\bf F} \rangle({\bf r},t)]^2 
  \right)~, 
\end{eqnarray} 
with $g_s \equiv 4\pi a_s \hbar^2/m$ the appropriate coupling 
constant for spin-density fluctuations. Using this action we can 
now investigate the effect of the above mentioned fluctuations by 
substituting $\zeta({\bf r},t) 
   = \zeta_{\rm AF}({\bf r},t) + \delta\zeta({\bf r},t)$ and 
expanding the action up to quadratic order in $\delta\zeta({\bf r},t)$. Solving then the Euler-Lagrange equation for 
$\delta\zeta({\bf r},t)$ and substituting the solution back into 
the action, we ultimately find the desired low-frequency result \cite{fei}
\begin{equation} 
\label{time} 
S_{\rm dyn}[n,\zeta] = \int dt \int d{\bf r}~ 
\frac{\hbar^2}{2g_s} 
    \left( \frac{\partial {\bf m}({\bf r},t)}{\partial t} 
    \right)^2~. 
\end{equation} 
While performing the calculation, we must make sure that we are 
not considering fluctuations of the spinor within the 
antiferromagnetic order parameter space. This requires the matrix 
elements $\zeta_{\rm AF}^{\dagger}({\bf r},t) {\bf F} 
\delta\zeta({\bf r},t)$ to be real, because for fluctuations within the antiferromagnetic parameter space we have in lowest order that
${\bf 0} = \langle {\bf F} \rangle ({\bf r},t) =
 \zeta_{\rm AF}^{\dagger}({\bf r},t) {\bf F} \delta\zeta({\bf r},t)
 + \delta\zeta^{\dagger}({\bf r},t) {\bf F} \zeta_{\rm AF}({\bf r},t)$.
Moreover, the normalization of 
the spinor requires also that $\zeta_{\rm AF}^{\dagger}({\bf r},t) 
\delta\zeta({\bf r},t) = 0$. 
 
The importance of this result is twofold. First, from 
Eqs.~(\ref{gradient}) and (\ref{time}) we see that at the quantum 
level the dynamics of the space-independent part of the vector 
field ${\bf m}({\bf r})$ is governed by the following 
time-dependent Schr\"odinger equation 
\begin{equation} 
i\hbar \frac{\partial}{\partial t} \Psi({\bf m},t) 
   = - \frac{\hbar^2}{2I} \mbox{\boldmath $\nabla$}_{\bf m}^2 \Psi({\bf m},t) 
\end{equation} 
for the wave function $\Psi({\bf m},t)$. It thus corresponds 
exactly to a quantum rotor with a moment of inertia equal to $I = 
\hbar^2 V_0(\mu)/g_s$, where $V_0(\mu)$ is the total volume of the 
spinor condensate in the Thomas-Fermi limit. In an harmonic trap the moment of inertia is thus proportional to the $3/5$ power of the total number of atoms.
The eigenstates of this Schr\"odinger equation are the spherical harmonics
$Y_{S,M_S}({\bf m})$. In this way we thus recover the fact that 
according to quantum mechanics both the total spin of the Bose-Einstein 
antiferromagnet as well as its projection on the quantization axis must always be an integer. More precisely, since the ground state wave function is given by $Y_{0,0}({\bf m}) = 1/\sqrt{4\pi}$, we have 
actually shown that the many-body wave function of the 
antiferromagnetic spinor condensate is a singlet state exactly 
\cite{nick,jason2}. Note that physically this phenomenon is equivalent to the 
way in which `diffusion' of the overall phase of a Bose-Einstein 
condensate leads to the conservation of particle number 
\cite{li,henk1}. The main difference is that here the `diffusion' takes place on the surface of a unit sphere instead of on a unit circle.
 
Second, and most important for our purposes, we can now determine the single monopole dynamics, by using the ansatz 
${\bf m}({\bf r},t)= {\bf m}_{\rm HP}({\bf r}-{\bf u}(t))$ for the texture of a moving monopole, which is 
expected to be accurate for small velocities $d{\bf u}(t)/dt$ and, 
in the inhomogeneous case, near the center of the trap where 
$u/R_{\rm TF} \ll 1$. Substituting this ansatz into 
Eq.~(\ref{time}) and remembering also Eq.~(\ref{pot}), we find that the 
action for the center-of-mass motion of the monopole becomes precisely that 
of a massive particle 
\begin{equation} 
S_{\rm HP}[{\bf u}] = \int dt~ \left( 
 \frac{m_{\rm HP}}{2} \left( \frac{d{\bf u}(t)}{dt} \right)^2 
 - E_{\rm HP}[{\bf u}] 
 \right)~, 
\end{equation} 
with a mass given by $m_{\rm HP} = 2mR_{\rm TF}/a_s$. The semiclassical equation of motion for the monopole position is therefore simply
Newton's equation
\begin{equation}
m_{\rm HP} \frac{d^2{\bf u}(t)}{dt^2} 
   = - \mbox{\boldmath $\nabla$}_{\bf u} E_{\rm HP}[{\bf u}]~.
\end{equation}
In view of the fact that the energy of the 't Hooft-Polyakov monopole decreases as the distance to the center of the trap increases, we conclude that in general the monopole is always accelerated to the boundary of the condensate. Typically
it will reach that boundary in a time interval of order 
\begin{equation} 
\Delta t_{\rm HP} 
  \simeq \frac{\pi}{2} \sqrt{\frac{m R_{\rm TF}^2}{2\mu} \frac{a_n}{a_s}} 
  = \frac{\pi}{2\omega} \sqrt{\frac{a_n}{a_s}}~, 
\end{equation} 
which for present-day experiments is sufficiently long to be able to observe the monopole once it is created \cite{klaus}. In this context it should be noted that all our calculations are performed at zero temperature. In the presence of a normal component, the monopole experiences damping, which increases the time needed to reach the edge of the spinor condensate. 

Summarizing, we have investigated the most important equilibrium and nonequilibrium properties of a 't Hooft-Polyakov monopole in a trapped, antiferromagnetic Bose-Einstein condensate. A further direction of research is an {\it ab initio} calculation of the above mentioned friction force on the monopole. In fact, we expect the thermal cloud to lead not only to dissipation but also to noise. Both effects can be conveniently treated within the general framework of the stochastic field theory that was developed previously for nonequilibrium  phenomena in partially Bose-Einstein condensed gases \cite{henk1}. Another interesting topic is the interaction between two monopoles and the many-body properties of a gas of these topological objects. Finally, a very important experimental problem is the creation of a 't Hooft-Polyakov monopole. Of course, pairs of monopoles with opposite winding numbers can in principle be created in a thermal quench or by sufficiently shaking up the spinor condensate. However, a more controled creation mechanism is desirable. Therefore, we are presently exploring if a single monopole can be created by an appropriate tailoring of the detuning, the intensity and the polarization of two pulsed Raman lasers.   
 
This work is supported by the Stichting voor Fundamenteel Onderzoek der 
Materie (FOM), which is financially supported by the Nederlandse 
Organisatie voor Wetenschappelijk Onderzoek (NWO).

\end{multicols} 

\begin{references}
\bibitem{spin} Note that we call the vector field
               ${\bf m}({\bf r})$ a spin texture even
               though its effect is only visible in the spin fluctuations,
               i.e., for example in 
               $\langle F_{\alpha}F_{\beta} \rangle({\bf r}) =
                \delta_{\alpha\beta} - m_{\alpha}({\bf r})m_{\beta}({\bf r})$.
\end{references}
\end{document}